\documentclass[fleqn,twoside]{article}
\usepackage{espcrc2}

\newcommand{\AmS}{{\protect\the\textfont2
  A\kern-.1667em\lower.5ex\hbox{M}\kern-.125emS}}

\newcommand{\beq}{\begin{equation}}
\newcommand{\eeq}{\end{equation}}
\newcommand{\be}{\begin{eqnarray}}
\newcommand{\ee}{\end{eqnarray}}

\def\+{\dagger}

\title{Dark Matter as  Dense Color Superconductor}
\author{Ariel R. Zhitnitsky
 \address{Department of Physics
 and Astronomy, University of British Columbia,
  Vancouver, BC V6T 1Z1, Canada}}

\begin{document}

\begin{abstract}
 We discuss a novel cold dark matter candidate which
 is formed from the ordinary quarks
during the QCD phase transition when  the  axion domain wall
undergoes an unchecked collapse
due to the tension in the wall. If a large number of quarks
is trapped inside the bulk of a  closed  axion domain wall, the collapse 
 stops due to the internal  Fermi pressure. In this case
the system in the bulk, may reach the critical 
density when it undergoes a phase transition to a color  superconducting phase
with the ground state  being the  quark condensate,
 similar to BCS theory. If this happens, the new state of matter
representing the diquark condensate with a 
large baryon number $B \geq 10^{20}$
  becomes  a stable soliton-like  configuration. Consequently, it may 
serve as a  novel cold dark matter candidate.  
\end{abstract}

\maketitle
\section{Introduction}
The presence of large amounts of non-luminous components in the Universe has been
known for a long time. In spite of the recent advances in the field\cite{snowmass}, the mystery of the 
dark matter/energy remains: we still do not know what it is. The main goal of this talk
is to argue that the dark matter could be  nothing but well-known quarks
which however are not in the ``normal" hadronic phase, but rather
in some ``exotic", the so-called color superconducting (CS) phase. 

This is a novel phase in QCD
when light quarks form the condensate in diquark channels, and it  is  analogous to Cooper pairs of 
electrons in ordinary superconductors described by BCS theory. There existence 
of CS phase in QCD represents our first crucial
element for our scenario to work. The study of CS phase received a lot of attention   
last few years \cite{cs_n}.
 It has been known that this regime may be realized in nature in neutron stars
interiors and in the violent events associated with collapse of massive stars or collisions 
of neutron stars, so it is important for astrophysics. The goal of this talk is to argue
that such conditions may occur in early universe during the QCD phase transition,
 so it might be important for cosmology as well.

 The force which squeezes quarks in neutron stars is gravity; the force which does
a similar job in early universe  during the QCD phase transition is 
a violent collapse  of a bubble formed from  the axion domain wall.
 If  number of quarks trapped inside  the bulk
 is sufficiently large, the collapse 
 stops due to the internal Fermi pressure. In this case the system
in the bulk  may reach the critical 
density when it undergoes a phase transition to CS phase 
with the ground state  being the  diquark condensate. 
We shall call the configuration with a large number of quarks in color superconducting phase 
formed during the QCD phase transition as the QCD-ball.
Therefore, an existence of the axion domain wall
represents our second crucial element for our scenario to work.
As is known, the axion
is one of the favorite candidates for the cold dark matter,
 see recent reviews \cite{axion_r}. 
In the present scenario the axion field plays the role of squeezer rather than 
dark matter itself.

 As  will be  shown below, once
such a configuration is formed, it will be extremely stable soliton like particle.
The source of the stability of the QCD-balls is related to the fact that
its mass $M_B$ growth as $M_B\sim B^{8/9}$ for the large baryon (quark) charge $B$
and becomes smaller than the mass of a collection of free separated nucleons with the same
baryon charge. The region of the absolute stability of the QCD-balls is determined by inequality
$m_N >\frac{\partial M_B}{\partial B}$ which can be always satisfied for very large $B >B_c$.
However, even for sufficiently large $B$ but smaller than $B_c$, i.e. $1\ll B< B_c$, we  still
expect that QCD-balls will be very long-lived metastable states. 
This is because the elementary excitations carrying 
the baryon charge in CS phase
and hadronic phase are very different: in CS phase they carry color charge along with baryon 
charge; in hadronic phase
they are color singlets. 
Therefore, if sufficiently large number of quarks   is trapped inside the 
axion bubble during its shrinking, it may result in formation of an absolutely stable 
( or long lived, metastable) QCD-ball with the ground state being a diquark condensate.  
 Such QCD-balls, therefore,  may serve as the cold dark matter candidate and 
contribute to $\Omega_{DM}\simeq 0.3 $.

 Strictly speaking, the QCD-balls being the baryonic configurations, would behave like 
nonbaryonic dark matter. In particular, QCD-balls, in spite of their QCD origin, would not 
contribute to  $\Omega_{B}h^2\simeq 0.02$ in nucleosynthesis calculations because 
the QCD-balls would complete the formation by the time 
when temperature reaches the relevant for nucleosynthesis
region   $T\sim 1 MeV$. Once QCD-balls are formed, their
 baryon charge  is accumulated in form of the diquark condensate, rather than in form
 of free baryons, and in such a form the baryon charge is not available for  nucleosynthesis.
\section{QCD-balls} 
Crucial for our scenario is the existence of a squeezer, axion domain wall which will be formed
 during the QCD phase transition. We assume that the standard problem of the 
domain wall dominance is resolved in  some way as discussed previously
in the literature \cite{axion_r}.
 We also assume that quarks which are trapped inside of the axion bubble, can not
 easily escape 
 the bulk when the bubble  is shrinking. In different words, the axion domain wall is not 
transparent due to the QCD sandwich structure of the wall  as 
 discussed in \cite{FZ},\cite{SG}. The collapse is halted due to the Fermi pressure.
 Therefore, we assume that  a large number of quarks remains in the bulk of volume $V$
 when  the Fermi pressure cancels the surface tension. Exactly this condition determines
a typical radius $R_0$ of the QCD-ball
 at the moment of formation. The pressure due to the surface tension is 
$P_{\sigma}=\frac{2\sigma}{R}$,
while the fermi pressure $P_f$ in the bulk can be easily estimated from the thermodynamical potential 
for the  non-interacting relativistic fermi gas, $\Omega=-P_{f}V$ with $\Omega$
given by:
\be  
\label{1}
\Omega=gV\int^{\mu}_0\frac{(p-\mu)d^3p}{(2\pi)^3}=-\frac{gV\mu^4}{24\pi^2}=
-P_{f}V,
\ee
Here $\mu$ is the Fermi momentum of the system  to be expressed in terms of the fixed
baryon charge $B$ trapped in the bubble; 
$$B= g V\int^{\mu}_0\frac{d^3p}{(2\pi)^3}= \frac{2g}{9\pi}\mu^3R^3,~~
\mu=\left(\frac{9\pi B}{2 g R^3}\right)^{\frac{1}{3}},$$
  $g\simeq 2N_cN_f =18$ is the degeneracy factor
 for massless degrees of freedom.
The condition $P_{\sigma}\simeq P_{f}$ determines a typical radius of the QCD-ball
with a fixed initial baryon charge $B$ at the moment of formation:
\be
\label{2}
P_{f}\simeq P_{\sigma} \Rightarrow~ \frac{g}{24\pi^2}
\left(\frac{9\pi B}{2 g R^3}\right)^{\frac{4}{3}}\simeq \frac{2\sigma}{R} 
~\Rightarrow~ \nonumber \\
R^3_0  \simeq  \frac{c B^{4/3}}{8 \pi \sigma} ,~~~~~~
c \equiv \frac{3}{4} (\frac{9\pi }{2 g} )^{1/3}\sim 0.7. ~~~~~~~
\ee
 Here $\sigma\simeq  f_am_{\pi}f_{\pi}$ is the axion domain wall tension with $f_a 
\sim (10^{10}-10^{12}) GeV$ being  constrained by the axion search experiments. 
As the first approximation, we neglected many contributions  in (\ref{1}, \ref{2}) 
which might be important numerically but can not drastically change  our main results.
In particular, we neglected 
the quark-quark interaction on 
the Fermi surface, which brings the system into  superconducting phase for relatively
 large  $\mu \geq 400 MeV$ \cite{cs_n}. The corresponding contribution gives some correction to the energy
 of order $(\frac{\Delta}{\mu})^2 $ with $\Delta\sim 100 MeV$ is the  superconducting gap,
 and can be neglected for the asymptotically large $\mu$.  
We also neglected the  so-called bag constant contribution which is a phenomenological
way to  simulate the confinement.
 Parametrically this  contribution is also suppressed as
 $(\frac{\Lambda_{QCD}}{\mu})^4$ but in reality might be important.
Let us emphasize: we are not attempting to solve the problem of equilibrium 
between CS and hadronic phases\footnote{ Such a problem has been 
recently discussed in ref.\cite{Krishna}
where  the interface region between 
nuclear matter and CFL (color flavor locking) superconducting phase was analyzed.}. Rather, 
we are attempting to estimate the initial size $R_0$ of the QCD ball which is formed
due to the collapse of the axion domain wall halted by the fermi pressure. 
Other contributions, such as quark-quark interaction 
 or the bag constant, may change the numerical estimates for $R_0$ however
they can not replace the main player of the game, the fermi pressure which is
responsible for the stopping of the collapse.  

Once the QCD ball with radius $R_0$ is formed, the bulk of this object being in CS phase
will be quite stable with respect to the decay on the free nucleons if some conditions are met, 
see below.
The stability remains intact even if the original axion domain wall would decay: the 
configuration remains  stable
due to the CS phase itself, and not because of the support from the axion domain wall. 
The only role
the axion domain wall plays in our scenario is the squeezing the matter at the very first 
instant
in order to bring the large number of quarks into the CS phase. The later stages of the 
evolution of the QCD ball
will determine the precise 
 structure of this   configuration at the equilibrium, including the interface region separating 
the CS phase with nuclear matter phase; we do not expect, though,
 that this evolution would considerably change our estimates
for $R_0$ presented above.
 
Once a typical radius $R_0$ for a QCD ball  is estimated (\ref{2}), 
  the total energy $E$ (  fermion energy + axion domain wall energy) and baryon number density 
$n$ of the QCD-ball   can be also estimated 
with the result 
\be
\label{3}
E \simeq  4 \pi \sigma R_0^2  + gV \frac{\mu^4}{8\pi^2} 
 \simeq \frac{3}{2}(8\pi c^2)^{\frac{1}{3}}B^{8/9}\sigma^{1/3}   \nonumber \\
  n = \frac{B}{V} \simeq \frac{6\sigma}{c}B^{-1/3} ~~~~~~~~~~~~~~~~~~~~~~~~~~~~~~~
\ee 
 The most important consequence of this result is the behavior $E\sim B^{8/9}$ which implies 
 that for sufficiently
large $B$ the QCD-ball becomes an absolutely stable object. It happens when $\frac{m_N}{3} > 
\frac{\partial E}{\partial B}$,
i.e.\footnote{Our normalization for the baryon charge corresponds to $B=1$ for the quark,
 thus factor $\frac{m_N}{3}$.}
\be
\label{4}
B > B_c \sim (2^9\pi c^2\frac{\sigma}{m_N^3})^3\simeq 10^{33}, ~ n_c\simeq 9 n_0,
\ee
where $n_0$ is the nuclear saturation density, $n_0\simeq 0.16 (fm)^{-3}$.
Numerically,   $ B_c \simeq 10^{33}$   corresponds to 
stabilization radius $R_c\simeq 10^{12} GeV^{-1} $ and mass $E_c\simeq 10^{33} GeV$.
Here we use $f_a\simeq 10^{10} GeV$ which corresponds to $\sigma\sim 1.8\cdot 10^8 GeV^3$.
As we mentioned in Introduction, we expect that the  QCD-balls with much
 smaller $B$ can live long enough to serve 
as a dark matter. 
 Therefore, in what follows we   use
 a more realistic value for $B_c^{exp}\sim 10^{20}$ which follows from   the
 experimental bounds on masses and fluxes of nontopological solitons, see below.

  Few remarks are in order. First,  we note that  there is an upper limit on value 
$\bar{B}$  above which the QCD-balls can not be formed at the very first instant during the 
collapse of the domain walls.
This  follows from the expression (\ref{3})  for the baryon number density   
  which decreases when B increase. Therefore, 
at some point, the density $n$ reaches the magnitude where the phase transition between nuclear
 matter and CS phase occurs. 
Numerically it happens  at  $\bar{B} \sim 10^2 B_c$. However, it does not preclude
from formation of  a larger object at the later stages of the QCD-ball evolution.

As our next remark, we would like to mention some similarity
between the QCD-balls (which is the subject of this letter) and Q-balls\cite{qball} which are 
 nontopological solitons 
associated with some conserved global $Q$ charge.  
In both cases, 
a soliton mass as function of $Q$  has behavior, similar  to our eq. (\ref{3}), and therefore, 
it may become a stable configuration for relatively large $Q$ charge. Therefore, an effective
 scalar field theory
with some specific constraint on potential (when
  $Q$ ball solution exists) is realized for QCD in high density regime.  
The big difference, of course, is that underlying theory for QCD-balls is well known: it is QCD
 with no free parameters, in huge contrast with the theory of   Q-balls. 
Formal similarity becomes even more striking if  one takes into account that the ground state 
of the
CFL phase in QCD  is determined by the diquark condensate with the following  time 
dependence  
$   \langle \Psi\Psi\rangle^* 
     \sim (e^{i2\mu t}),  
$
with $\Psi$ being the original QCD quark fields, and $\mu$ being the chemical potential of the 
system,
see formula (40) from ref. \cite{FZ2}.
As is known, such time-dependent phase is the starting point in construction of the 
Q balls\cite{qball}.
  
Finally, we want to constraint the QCD-ball parameters
  from available data. 
We follow \cite{qball_exp}  and assume that a typical cross section of a neutral QCD-ball
  with matter is determined by their geometrical size, $\pi R_0^2$.
  We also assume that the QCD-balls is  the main 
contributor  toward the dark matter in the Galaxy,
 in which case their flux $F$  should satisfy
\be
\label{dark}
F <  \frac{\rho_{DM} v}{4\pi M_B} \sim 
7.2\cdot 10^5 \frac{ GeV}{M_B} cm^{-2}sec^{-1}sr^{-1},
\ee
where $\rho_{DM}$ is the energy density of the dark matter in the Galaxy, 
$\rho_{DM}\simeq \frac{0.3 GeV}{cm^3}$, and $v\sim 3\cdot 10^{-3}c$ is
 the Virial velocity of the QCD-ball.
We identify $M_B$ in the expression (\ref{dark}) 
with the total energy $E$ (\ref{3}) of the QCD ball at rest with given baryon  charge $B$.
Different experiments discussed in \cite{qball_exp} 
give the following bound on the flux of neutral soliton-like objects:
$
F~ <~ const. \cdot 10 ^{-16} cm^{-2}sec^{-1}sr^{-1},
$
This translates to the following 
lower limit of the neutral QCD-ball mass $M_B$ and baryon charge $B$,
\be
\label{9}
M_B^{exp} ~ > ~ 2\cdot 10^{21}~ GeV ,~~~ 
B^{exp}  ~ >~ 1.6\cdot 10^{20}.
\ee
These experimental bounds are well below the critical line of the {\it absolute} stability
of the QCD-balls, given by eq. (\ref{4}). However, as we mentioned above, 
we expect that QCD-balls will be very long-lived metastable states even if their baryon
charge is below the critical limit (\ref{4}). Therefore, we 
consider experimental constraint  (\ref{9}), 
rather than a theoretical stability bound (\ref{4}) 
as a more realistic bound on $M_B,~B$ for the QCD-balls.
  
We have seen that   
 the observed relation $\Omega_B\sim\Omega_{DM}$ within an order of magnitude 
finds its natural explanation in this scenario: both contributions are  originated from the 
same physics at the same instant during the QCD phase transition. As is known, this fact is extremely
difficult to explain in models that invoke a dark matter candidate not related to baryons.
We have also seen that
without adjusting any  parameters the baryon density inside the QCD-ball
falls exactly into  appropriate region
 of the QCD phase diagram where color superconductivity
takes place.
The scenario is no doubt lead to important consequences for cosmology and astrophysics,
which are not  explored yet.
In particular,  some unexplained events, such as Centauro events, or  even the Tunguska-like
events (when no fragments or chemical traces have ever been recovered),
can be related to the very dense QCD balls.
 If this is the case, the arrival directions
should correlate with the dark matter distribution and show the halo asymmetry.
Also: recent discovery\cite{NASA}  suggests
that the matter in some stars could be  even denser than nuclear
matter. It could be also   related to the very dense QCD balls.
 Therefore, the ``exotic", dense color superconducting phase in QCD, might be much  
more common state of matter in the Universe  than the ``normal" hadronic phase we know.
In conclusion, qualitative as our arguments are, they suggest that 
 the dark matter   could be  originated at the QCD scale.

\end{document}